\documentclass[sigconf]{acmart}

\acmConference[Arxiv]{}{2023}{}

\copyrightyear{2023}
\acmYear{2023}
\setcopyright{rightsretained} 
\AtBeginDocument{%
  \providecommand\BibTeX{{%
    Bib\TeX}}}

\usepackage{algorithmic}
\usepackage{graphicx}
\usepackage{textcomp}
\usepackage{graphicx}
\usepackage{colortbl}
\usepackage{etoolbox}
\usepackage{multirow}
\usepackage{makecell}
\usepackage{color}
\usepackage[utf8]{inputenc}
\usepackage{xspace}
\usepackage{colortbl}
\usepackage{pifont}
\usepackage{tabularx}

\def\BibTeX{{\rm B\kern-.05em{\sc i\kern-.025em b}\kern-.08em
    T\kern-.1667em\lower.7ex\hbox{E}\kern-.125emX}}
    
\begin{document}

\title{Scratch Copilot Evaluation: Assessing AI-Assisted Creative Coding for Families}

\author{Stefania Druga}
\affiliation{%
  \institution{Information School, University of Washington}
  \city{Seattle}
  \state{Washington}
  \country{United States}}
\email{st3f@uw.edu}

\author{Nancy Otero}
\affiliation{%
  \institution{Kitco}
  \city{San Francisco}
  \state{California}
  \country{United States}}
\email{nancy.otero.o@gmail.com}

\renewcommand{\shortauthors}{Druga et al.}

\begin{teaserfigure}
\centering
\includegraphics[width=5.5in]{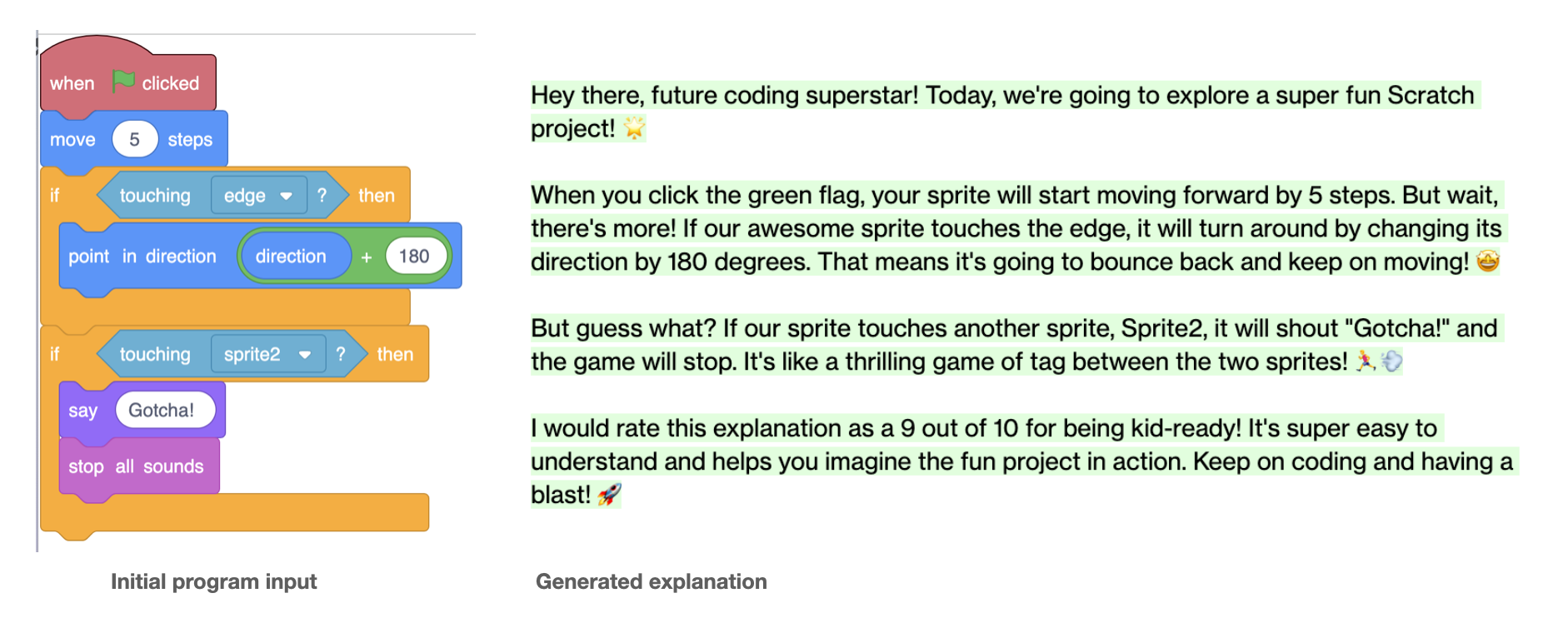}
\caption{Examples of code explanation provided by the GPT4 model.}
\label{fig:code_explain}
\end{teaserfigure}

\begin{abstract}
How can AI enhance creative coding experiences for families? This study explores the potential of large language models (LLMs) in helping families with creative coding using Scratch. Based on our previous user study involving a prototype AI assistant, we devised three evaluation scenarios to determine if LLMs could help families comprehend game code, debug programs, and generate new ideas for future projects. We utilized 22 Scratch projects for each scenario and generated responses from LLMs with and without practice tasks, resulting in 120 creative coding support scenario datasets. In addition, the authors independently evaluated their precision, pedagogical value, and age-appropriate language. Our findings show that LLMs achieved an overall success rate of more than 80\% on the different tasks and evaluation criteria.
This research offers valuable information on using LLMs for creative family coding and presents design guidelines for future AI-supported coding applications. Our evaluation framework, together with our labeled evaluation data, is publicly available \footnote{https://github.com/stefania11/ScratchCopilot-Evaluation}.
\end{abstract}


\maketitle

\section{Introduction}

Computer Science (CS) education faces a critical bottleneck. The need for more trained teachers and curriculum designers stifles progress in this field \cite{yadav2016expanding}. Supporting project-based learning for youth and families, particularly creative coding, could be a potential solution. Prior work shows that engaging youth and families in creative coding has been advocated to promote more inclusive and accessible learning experiences \cite{roque2016family}. The limited availability of support and expertise in computer science education also calls for innovative technological solutions, similar to Github Copilot, which show considerable promise \cite{imai2022github}.

Parents often need more technical knowledge for effective coding instruction despite their expertise in engaging their children. In this regard, models have been proposed to complement joint creative coding between children and parents by providing timely suggestions, questions, ideas, and tips. It is worth noting, however, that the introduction of external support can both positively and negatively impact youth motivation and learning.

Studies have shown that creative coding can significantly enhance student motivation and bolster confidence in their knowledge and technical abilities compared to traditional CS programs (Rittenhouse, C. S. Scholarship, Research, and Creative Work at Bryn Mawr College). Furthermore, creative coding might allow students to develop a more immersive and experiential relationship with digital processes, providing them with hands-on experiences and theoretical frameworks \cite{dufva2021creative}.

Young people have lauded experiences that enable them to express their ideas, foster relationships, assist others, and discover new perspectives about themselves. This emphasis extends beyond the common focus of coding initiatives on computational thinking and problem-solving skills to support social, leadership, and identity development \cite{roque2019youth}.

Large Language Models (LLMs), such as OpenAI’s Codex and GPT-3, have demonstrated potential in aiding tasks related to explaining, ideating, and debugging creative coding projects. However, their current performance may fall short of fulfilling the unique needs of middle school families \cite{pearce2022s}. While these models have achieved some success in generating novel and meaningful content, the necessity for human oversight to ensure the quality and accuracy of the generated content remains \cite{sarsa2022automatic, leinonen2023using}.

Despite the potential of such tools, several potential disadvantages exist when utilized for family creative coding. For instance, young learners may need to be more responsive to these tools, impairing their ability to create similar code independently. Other challenges include formulating their intentions to generate the desired code and understanding the code produced by AI for subsequent modification if needed \cite{vaithilingam2022expectation}. Moreover, a recent study assessing a new creative coding integrated development environment (IDE) revealed students’ concerns about the trade-off between improving their abilities and facilitating the development of their skills through IDE syntax templates and autocomplete coding features \cite{mcnutt2023study}.

Involving parents as learning partners in the creative process is paramount. Past research has shown that parents can act as mentors and co-tinkerers when families engage in game programming \cite{druga2022families} or AI literacies tinkering \cite{Druga2022Family,long2022family}.

Given these findings, our research aims to answer the following question:

\begin{itemize}
\item RQ: How well do large-language models support explaining, ideating, and debugging Scratch projects for middle school families?
\end{itemize}

This paper explores the potential of LLMs in aiding families interested in learning creative coding together. We focus on the applicability of LLMs for generating Scratch program explanations, debugging, and ideation support. Our previous user study identified these three areas as primary needs for family AI-assisted creative coding \cite{Druga2023DesignAI}.  

Our findings reveal that LLMs achieved an overall success rate of over 80\% across the various tasks and evaluation criteria. This study further contributes a public dataset of Scratch programs, complemented by the code explanations, debugging, and ideation support tasks we employed for the LLM evaluation.

Based on these findings, we discuss potential scenarios for designing inclusive LLM support for family creative coding. Moreover, we propose a series of design guidelines that could inform the development of future AI-supported coding applications. This exploration thereby provides insights into the effectiveness and potential of LLMs as supportive tools for families engaged in creative coding, offering a promising avenue for inclusive and accessible computer science education. Our evaluation framework, together with our labeled evaluation data, is publicly available here: github.com/stefania11/ScratchCopilot-Evaluation.

\section{Related work}

\subsection{LLMs in Computing Education}
Large language models (LLMs) have demonstrated potential in numerous fields, especially education and programming. In addition, the influence of LLMs on novice learners, particularly in introductory programming environments, has garnered scholarly interest.

Kazemitabaar et al. studied the effects of OpenAI Codex on middle school learners within a self-paced learning setting. Their findings suggest that Codex significantly enhanced code-authoring performance without negatively impacting manual code-modification tasks \cite{Kazemitabaar2023Studying}. However, it was observed that performance differences in post-tests conducted a week later were not statistically significant, underscoring the necessity of further research.

Leinonen et al. explored the use of LLMs in generating code explanations, comparing GPT-3-generated explanations with those created by students in an introductory programming course. Their findings highlighted that LLM-generated explanations were perceived as significantly easier to comprehend and more accurate than those produced by the students \cite{leinonen2023using}.

Turning to debugging tasks, Chen et al. introduced the concept of Self-Debugging, which trains LLMs to debug their predicted programs using few-shot demonstrations. Their study established that Self-Debugging surpassed performance standards on code generation benchmarks, improving the baseline accuracy by up to 12\% and demonstrating notable sample efficiency \cite{chen2023teaching}. Similarly, Madaan et al. examined LLMs’ capacity to suggest performance-improving code edits, establishing that tools like CODEGEN and CODEX could generate such edits for C++ and Python programs \cite{madaan2023learning}.

The potential of LLMs extends to computer science education, where AI code generators can offer substantial support to learners and educators alike. For example, they can automatically rectify semantic bugs and syntax errors, allowing learners to concentrate more on theoretical and problem-solving aspects of computational thinking. Additionally, these tools can assist educators in developing curriculum by creating programmatic exercises and explaining solutions \cite{sarsa2022automatic}.

Guo’s study offers further insight, introducing an interactive web-based tool, the online Python Tutor, which aids students in understanding Python programming through visualization of code execution \cite{guo2013online}. This supports novice learners in comprehending complex computer programming concepts and is an effective debugging aid.

In the realm of creative coding, it has been observed that media arts-related coding education attracts a diverse range of students who might not otherwise engage with CS in a formal setting \cite{greenberg2007processing,malita2020drawing,wood2016computational}. Studies such as those by Sáez-López 2016, MacNeil 2022, and Sarsa 2022 indicate the potential of LLMs in this sphere, particularly for middle school families \cite{sarsa2022automatic}. Furthermore, they demonstrate that LLMs, like GPT-3 and OpenAI Codex, can generate helpful code explanations and programming exercises, providing potential value in creative coding projects.

The current work highlights the potential of Large Language Models (LLMs) in various aspects of coding education, including enhancing code-authoring performance, generating understandable code explanations, and assisting in debugging tasks. Moreover, the promising role of AI code generators and AI-enhanced visualization tools in supporting learners and educators in computer science education has been underscored. However, despite these advancements, a gap persists in understanding LLMs’ effectiveness and potential limitations, particularly concerning youth and families engaged in creative coding.

This study seeks to fill this gap by investigating the utility of LLMs in the context of middle school families engaging in creative coding. We focus on the potential of LLMs for generating Scratch program explanations, debugging, and ideation support. In doing so, we aim to contribute to the growing body of research on using LLMs in coding education and provide valuable insights into their utility for this demographic.

\subsection{Family Creative Coding}
Creative coding, distinct from traditional Computer Science (CS) education, often adopts a \textit{bricolage approach} \cite{mclean2012computer}. This concept, introduced in the programming context by Turkle and Papert \cite{turkle1990epistemological}, paints the coder as a bricoleur, akin to a painter contemplating their canvas between brushstrokes. This approach casts programming as a collaborative venture with the machine, a conversation rather than a monologue, wherein mistakes are not missteps but opportunities for navigation and mid-course corrections.

However, it is crucial to note that the benefits of creative coding extend beyond motivation; they also challenge the assumption that creative coding inherently develops computational thinking and problem-solving skills. For example, a recent study found that novice students often need help using optimal strategies to create animations, even with explicit instruction \cite{woo2022problem}. This finding highlights the importance of pedagogical approaches in promoting computational thinking in the context of creative coding.

A growing body of research supports the idea that collaborative creative coding, mainly when supported by AI, can effectively engage both children and parents in learning and creating with technology. Prior studies have detailed successful programs where families participate in creative coding workshops, sparking interest and activity in computing among parents and children alike \cite{roque2016family, Bresnihan2019OurKidsCode}.

For instance, Druga et al. delved into how parents can aid their children in developing AI literacies through learning activities, emphasizing the benefits of parent-child partnerships \cite{Druga2022Family}. Another study by Zhang et al. introduced StoryBuddy, an AI-enabled system designed for parents and children to create interactive storytelling experiences. This system caters to dynamic user needs and supports various assessment and educational goals \cite{Zhang2022StoryBuddy}.

In summary, collaborative creative coding, bolstered by AI, can be a significant avenue to involve children and parents in learning and technological creation. The existing literature provides different approaches to designing and implementing such programs, informing our current study. Our research extends this work by focusing on the potential of Large Language Models (LLMs) to support such collaborative creative coding experiences, particularly for middle school families. This emphasis on LLMs in the family creative coding context adds a new dimension to the existing discourse, potentially expanding and enhancing these collaborative learning experiences.

\subsection{Culturally-Responsive Computing Education}
The advent of large language models (LLMs), capable of human-like language generation, has ushered in a new era of technological interaction, which can shape user behavior and opinions. Jakesch \cite{jakesch2023co} suggests that when LLMs express certain viewpoints more frequently than others, they may inadvertently influence user perspectives. The potential bias in LLMs is further substantiated by studies such as those by Gaci \cite{Gaci2022Masked} and Nadeem \cite{Nadeem2021StereoSet}, which report that pre-trained LLMs often reflect and perpetuate societal stereotypes and biases.

Addressing this concern, several studies have proposed measures to mitigate social biases in LLMs. For instance, Liang et al. \cite{PaulTowards} introduced new benchmarks and metrics for identifying and reducing these biases. In contrast, Mattern et al. \cite{Mattern2022Understanding} proposed a robust framework for quantifying biases exhibited by LLMs. Thus, it becomes crucial to confront and alleviate these biases, especially within the context of computing education.

Culturally Responsive Computing Education (CRC) is an evolving field emphasizing integrating students’ identities and experiences into learning. This approach, as championed by Solyst \cite{Solyst2022Insights} and Morales-Chicas \cite{Morales2019Computing}, is recognized as key to fostering equity and justice in K-12 education. Moreover, Solyst et al. underline the challenges in fostering a sense of connectedness in online CRC programs and propose strategies to address them \cite{Solyst2022Insights}. Araujo \cite{Arawjo2021Intercultural} further advocates for an intercultural approach, promoting relationship-building across differences.

The importance of culturally-responsive approaches is also underscored in the context of AI education for K-12 students \cite{Eguchi2021Contextualizing}. These approaches include personalizing the learning experience, promoting AI ethics understanding among middle school students, and using cultural artifacts to reinforce computing concepts \cite{Anohah2020Conceptual}. Moreover, the role of collaborative engagement between schools and communities in fostering equity-oriented CS education is highlighted \cite{Lachney2021Teaching}.

In summary, the research underscores the significance of culturally-responsive approaches in computing and AI education and the necessity to address and mitigate the potential biases in LLMs. This current study aims to extend this discourse by exploring the potential of LLMs in a culturally-responsive, family-based creative coding context. Furthermore, we aim to contribute to the ongoing discussion on how to best leverage these advanced tools in a way that respects and integrates diverse cultural perspectives, ultimately promoting an inclusive and effective computing education.

\section{Method} 

\subsection{Development and Analysis of Scratch Projects}
\begin{figure*}[t]
\centering
\includegraphics[width=5.5in]{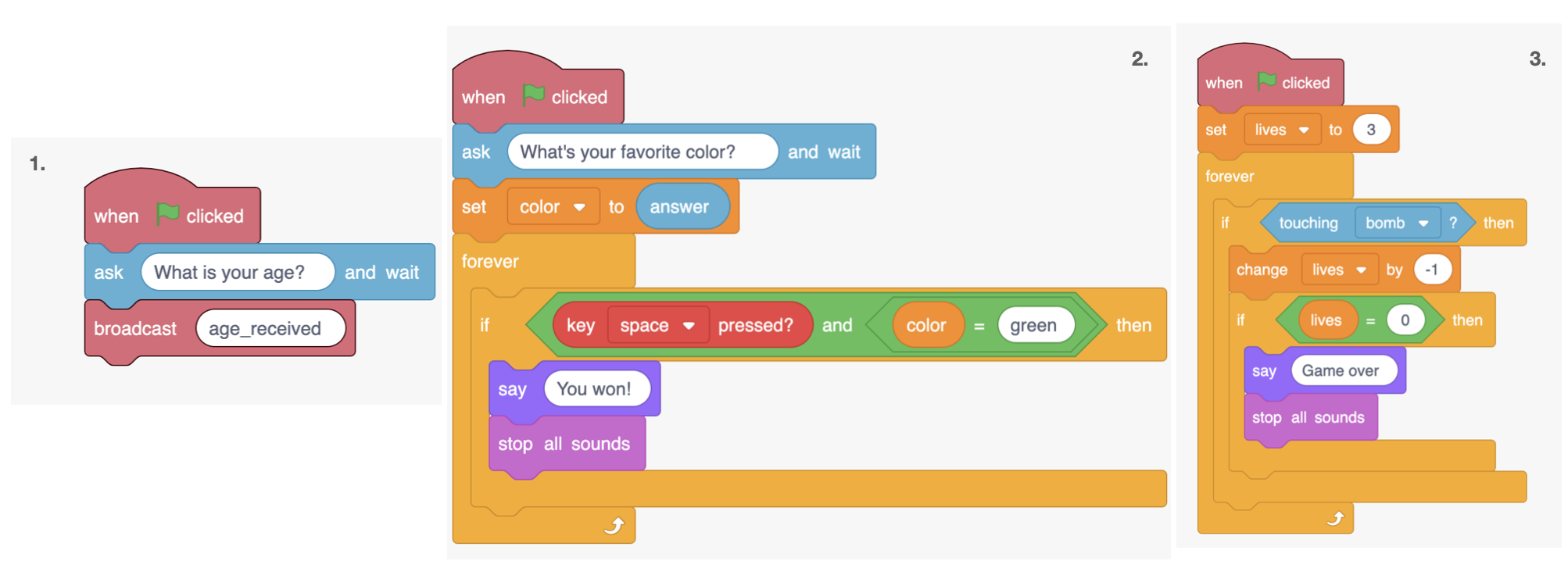}
\caption{Examples of input Scratch programs provided as input to the LLM:1.Asking player age, 2. Asking the player to guess a favorite color, 3. Game for avoiding bombs}
\label{fig:input_programs}
\end{figure*}

In our study, we curated a collection of 22 Scratch projects (see examples in Figure \ref{fig:input_programs}). These projects were selected by referencing popular Scratch community projects and salient examples from a previous study that analyzed 250,000 projects from the Scratch public repository \cite{aivaloglou2016kids}. These projects primarily aimed to evaluate the capabilities of a language learning model (LLM) in supporting code explanation, code debugging, and code ideation. In addition, these three tasks were identified as critical areas of focus based on findings from our previous user study on AI assistants for family creative coding. 

We utilized the 22 Scratch projects as inputs for an LLM (GPT4), generating responses with and without practice tasks. This resulted in a pool of 120 creative coding support scenarios. These scenarios were evaluated independently by the first two authors, focusing on precision, pedagogical value, and age-appropriate language. In cases where the two authors disagreed on the evaluation, they engaged in a discussion until a consensus was reached. Our evaluation framework, together with our labeled evaluation data, is publicly available \footnote{https://github.com/stefania11/ScratchCopilot-Evaluation}.

\subsection{Leveraging OpenAI's GPT4 Model}

OpenAI’s GPT-4, much like its predecessor GPT-3, can be interacted with through an API or a web interface. Users provide GPT-4 with a prompt, which the model uses as a foundation to generate content in alignment with the input. For example, upon receiving a natural language description of desired functionality, GPT-4 often generates corresponding source code.

We can guide the model’s content generation by specifying a “stop sequence” to halt generation upon reaching a particular sequence. Our study also employed options such as maximum token count for controlling the content length and “temperature” for influencing the model’s level of creativity or randomness. Lower temperature values decrease randomness by reducing the likelihood of generating less probable tokens. However, the model remains non-deterministic regardless of temperature, with variations in content across different runs, especially at higher temperature values.

After conducting several experiments, we settled on the following model parameters for our final evaluation: a temperature of 0.7, a maximum token count of 1024, and a maximum penalty P of 1. These parameters yielded the best results during our prompts testing.

The provided prompt significantly influences the generated content of GPT-4. Therefore, we prompted the model with an existing Scratch program and context-specific natural language instructions to guide GPT-4 in explaining, debugging, or ideating Scratch programs. For example, our prompts would reference the Scratch input program, ask the model to explain it to a middle school child, and evaluate its answer: 
\begin{quote}
    “You are an expert in creative coding for kids in middle school. Explain the following Scratch project \{scratch\_code\} in an accessible and fun way. Provide first a global overview of the project. \\ Rate your global response and show a score for how kid-ready your response was.” (prompt used for code explanation task without practice)
\end{quote}
 The complete list of prompts for creating our evaluation dataset is in the appendix.

\section{Evaluation} 
In the following section, we present the main findings of our LLM evaluation organized in the following creative-coding support scenarios: code explanation, debugging support, and ideation support. 
\begin{table*}[]
\centering
\begin{tabular}{cll}
\textbf{Coding Task} & \multicolumn{1}{c}{\textbf{\begin{tabular}[c]{@{}c@{}}Correct \\ suggestions\end{tabular}}} & \cellcolor[HTML]{FFFFFF}\textbf{Notes} \\ \hline
\rowcolor[HTML]{EFEFEF} 
\textit{Explain code} & 100\% & \begin{tabular}[c]{@{}l@{}}No errors, but the tone of language was overly enthusiastic \\ at times (i.e., “Hey there, Star coder,” “Super coder let’s look\\ at this program”)\end{tabular} \\
\textit{\begin{tabular}[c]{@{}c@{}}Explain code\\ with learning\end{tabular}} & 100\% & More line by line explanations in this mode. \\
\rowcolor[HTML]{EFEFEF} 
\textit{Debug code} & 80\% & \begin{tabular}[c]{@{}l@{}}The model would find bugs even in correct \\ programs, in two examples, it suggests creating \\ variables instead of finding conditionals errors\end{tabular} \\
\textit{\begin{tabular}[c]{@{}c@{}}Debug code \\ with learning\end{tabular}} & 90\% & \begin{tabular}[c]{@{}l@{}}Did not detect ``set score'' instead of  ``change score'' \\ and play concurrent sounds bugs\end{tabular} \\
\rowcolor[HTML]{EFEFEF} 
\textit{Code ideas} & 100\% & \begin{tabular}[c]{@{}l@{}}When the initial program was generic, it would\\ suggest similar suggestions such as adding counters,\\ timers, levels, multimodality\end{tabular} \\
\textit{\begin{tabular}[c]{@{}c@{}}Code ideas\\ with learning\end{tabular}} & 100\% & \begin{tabular}[c]{@{}l@{}}Whenever the initial program had something more specific\\  (i.e., reference to favorite color) the model would develop \\ that in creative ways otherwise, it would default to suggesting\\ common game mechanics prevalent in existing scratch projects\end{tabular}
\end{tabular}
\caption{summary of model evaluation}
\label{scratch_copilot_eval}
\end{table*}

\subsection{Code Explanation Evaluation}
When evaluating the code explanations, we studied whether all parts of the code were explained and whether each line was correctly explained. Of the 40 code explanations, 90\% explained all parts of the code (see Table \ref{scratch_copilot_eval}).

The LLM generated explanations to help middle schoolers understand their Scratch projects. In addition, TheLLMassistant provided engaging, age-appropriate explanations that were easy to understand, making the Scratch projects more enjoyable for the young coders. For instance, in the example illustrated in Figure \ref{fig:code_explain}, the LLM explained the project as “a thrilling game of tag between the two sprites,” making it relatable and exciting for the students.

The LLM successfully explained the Scratch code in various projects, breaking down each block’s steps and purpose in a way middle schoolers could comprehend. In other instances, the LLM explained a game where the character collects coins and avoids obstacles, detailing the code’s structure and the logic behind each section. This helped students grasp the concepts more effectively.

Although the LLM provided accurate explanations, the tone of language was sometimes overly enthusiastic, using phrases like “Hey there, Star coder” or “Super coder, let’s look at this program.” 

In several instances, the model explained more complex games where users collect coins and avoid obstacles to increase their scores. The model broke down each section of the code, explaining the start of the game, the continuous loop, the conditions when the sprite touches a coin or an obstacle, and the game’s conclusion when the score exceeds 100. It provides a systematic and detailed breakdown, clearly understanding each function.

In other examples, the LLM created appropriate metaphors for explaining more complex computational concepts such as variables or loops. For example, in one instance where the sprite’s size was changing continuously, the model likened the sprite to a balloon that inflates and deflates, making the code’s dynamics easier to understand. It explains how the sprite grows until it reaches a specific size, then starts to shrink, creating a continuous cycle. The model’s imaginative and engaging language makes the code’s purpose clear and appealing, encouraging students to explore further.

Overall, the LLM demonstrated its ability to support middle schoolers in understanding their Scratch projects through engaging, informative explanations.

\subsection{Code Debugging Evaluation}
\begin{figure*}[t]
\centering
\includegraphics[width=6.5in]{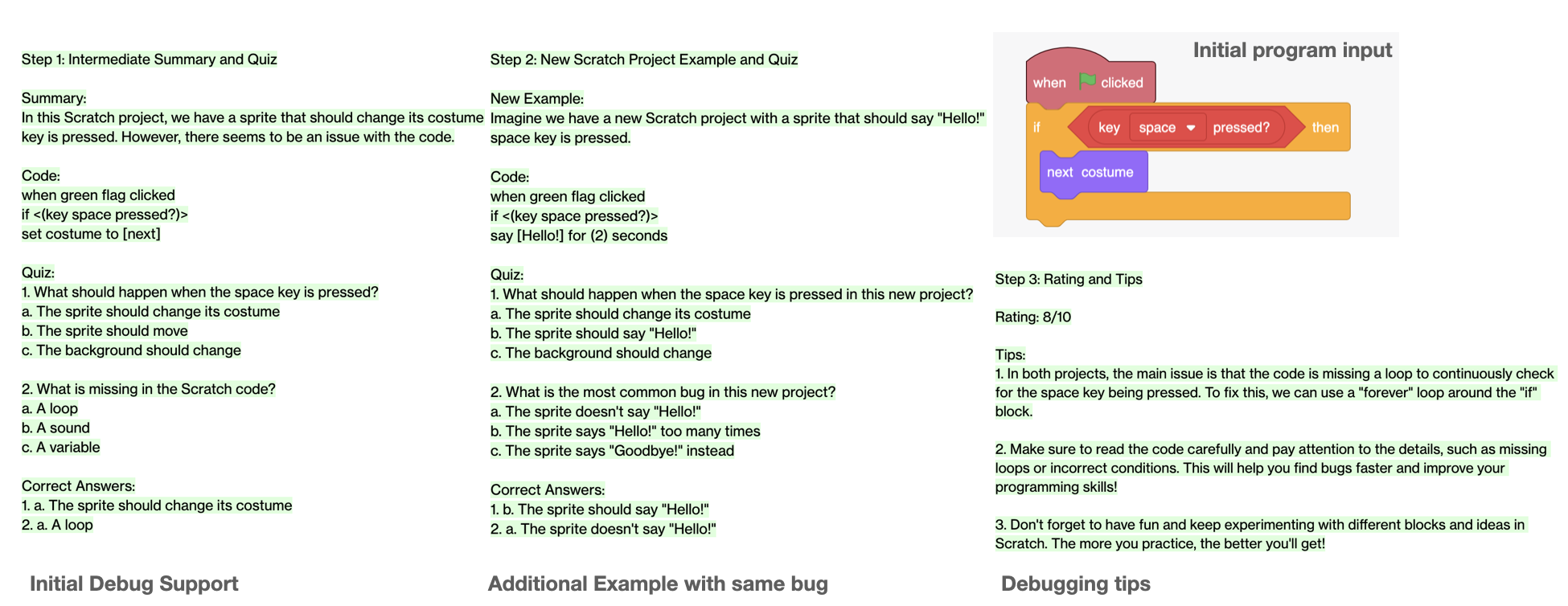}
\caption{Examples of debugging code support provided by the model.}
\label{fig:code_debug}
\end{figure*}

When evaluating the code debugging, we studied whether GPT4 could identify the correct bug and provide adequate support. Of the 40 code debugging examples, 80\% correctly identified the introduced bugs.

The LLM showed promise in aiding middle schoolers to debug their Scratch projects. However, although the LLM correctly identified bugs 80\% of the time, it occasionally suggested creating variables when the actual issue was related to conditionals.
These debugging errors were primarily due to the lack of context for the Scratch project input, as the model did not have access to prior code explorations or code edits from the same project. 

A noteworthy example of the debugging support proposed by the model involved the model using an interesting superhero analogy to elucidate the necessity of a “forever” loop in a student’s code, thereby enhancing the functionality of the project (see Figure\ref{fig:code_debug}).

In another scenario, the LLM identified that a student had not created a “lives” variable, confusing their project. The model guided the student through creating this variable, thus resolving the problem and making the project work as intended. However, the model did have some shortcomings; for instance, it failed to detect bugs, such as using “set score” instead of “change score” and suggested adding a “play sound” block rather than addressing the actual issue.

In the provided examples, the LLM generated interactive quizzes and practical tips as part of its pedagogical approach to facilitating debugging in Scratch projects. The quizzes were structured to summarize the project and then ask targeted questions about the intended behavior and the potential bug in the code. This approach emphasizes the comprehension of the code and encourages the learners to think critically about potential issues. For instance, in the first example, the LLM introduced a quiz highlighting the need for a loop to continuously check the space key press event (see Figure\ref{fig:code_debug}). 

Furthermore, the model provided tips after each interactive quiz session, offering advice on improving coding skills and debugging more effectively. These tips ranged from technical guidance, such as adding necessary loops or paying attention to missing elements in the code, to fostering a positive learning mindset, such as encouraging experimentation and maintaining a fun approach to coding. For example, in the second example, the model advised learners to provide clear instructions, experiment with different code blocks, and use simple language and engaging analogies. Combined with the quizzes, these insights establish an effective learning environment that fosters understanding and creativity in the coding process.

\subsection{Code Ideation Evaluation}
\begin{figure*}[t]
\centering
\includegraphics[width=5in]{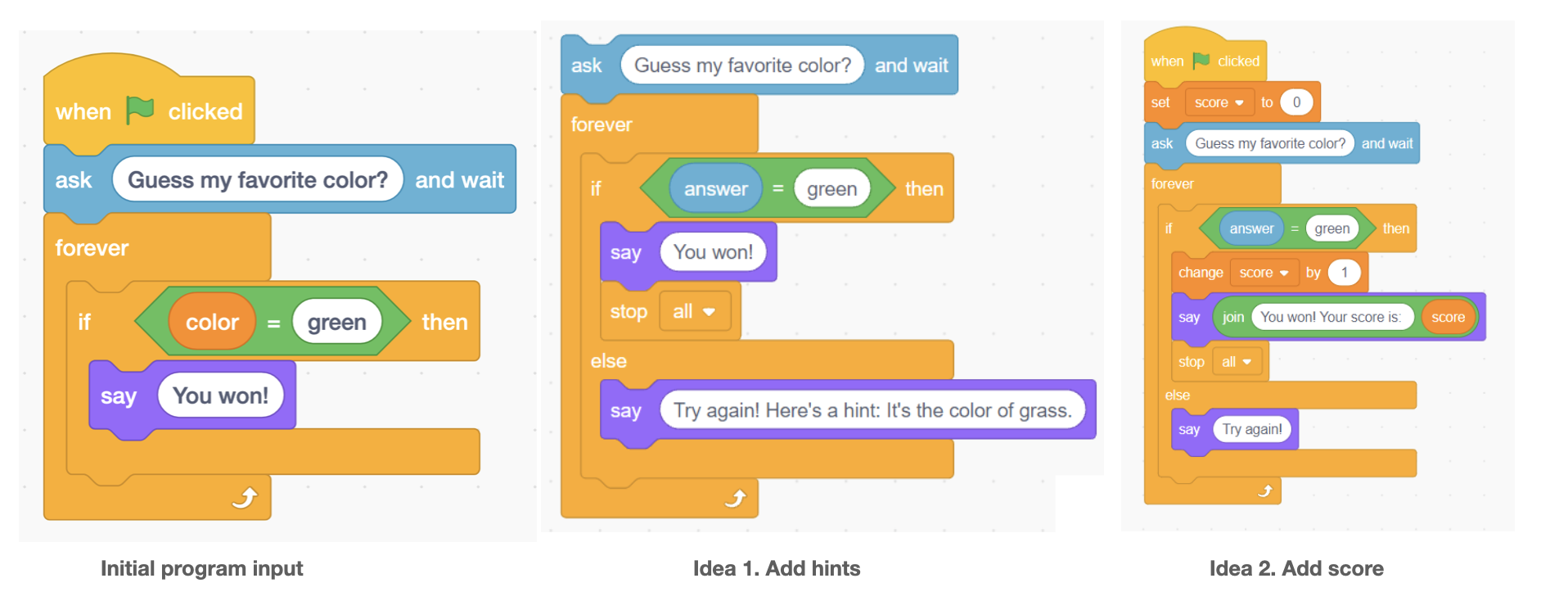}
\includegraphics[width=5in]{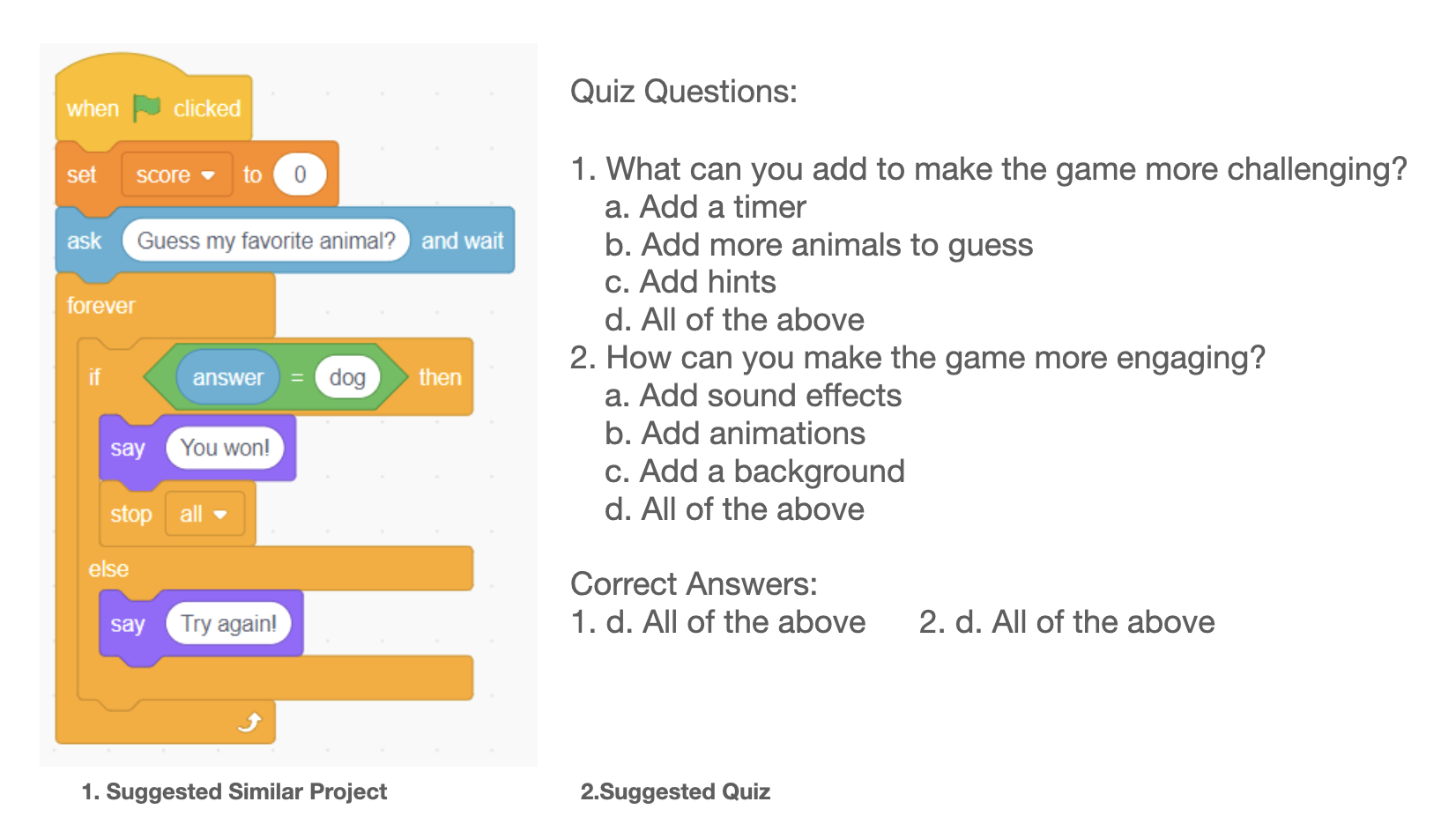}
\caption{Examples of code ideas provided by the model.}
\label{fig:code_ideas}
\end{figure*}
We studied whether LLMs could suggest relevant ideas for Scratch projects when evaluating the code ideations. All the suggestions made by the model were correct and written in kid-friendly language.

The LLM model suggested various ways to enhance Scratch projects for middle schoolers, focusing on interactivity, challenges, and visual appeal. In one example, the model proposed adding a sprite that follows the mouse pointer, introducing a timer for a time-based challenge, and altering the sprite’s costume to increase visual appeal. Additionally, the model recommended incorporating levels with increasing difficulty and sound effects and music to create a more engaging and entertaining experience.

Another example showcased the AI’s ability to offer creative suggestions for code modification, such as changing the sprite’s appearance and color with each loop iteration, creating a visually pleasing effect. The model also suggested incorporating user input to control the number of loop repetitions, making the project more interactive (see Figure \ref{fig:code_ideas}). Furthermore, the LLM proposed transforming the project into a game by adding collectible objects, a scoring system, and advancing levels with increasing difficulty.

In other instances, the LLM model focused on making the game more engaging and educational by adding hints based on color guesses. For instance, if a child guesses red, the hint informs them that their favorite color is of a cooler tone than red. This approach makes the game more engaging and promotes learning through feedback. The model also recommended adding a timer and associating sound effects with user actions to enhance the gaming experience (see Figure \ref{fig:code_ideas}).

\section{Discussion}
Our work asked: \textit {How well do large-language models support explaining, ideating, and debugging Scratch projects for middle-school families?}
Our study revealed that in the case of simple Scratch programs, LLMs such as GPT4 can achieve high precision and accuracy when generating code explanations, debugging supports, and code ideas. Moreover, we found that despite the model being overly enthusiastic sometimes, the language used in the support scenarios generated was child appropriate.

Our findings illuminate several implications for applying LLM models in supporting children’s creative coding. For instance, the model’s capacity to introduce new ideas and modifications often assumes a certain level of knowledge among children. Although this can benefit those with some experience, it may create difficulties for beginners. Therefore, future models must be designed with mechanisms to assess a child’s knowledge level and adjust their support accordingly. Also, these models should be equipped to identify and rectify any incorrect assumptions about the child’s knowledge or skill level.

Despite the LLM’s ability to generate suggestions, these were sometimes off-topic or made assumptions about Scratch’s capabilities that did not align with the child’s project goals. In addition, as in previous studies \cite{aivaloglou2016kids, robles2017software}, the LLM sometimes prompted children to write complex code, resulting in “code smells” or bad programming practices. This suggests that future LLMs should aim to restrict overly complex or off-topic suggestions, thereby providing more personalized and accurate support.

Our findings also pointed to the need for LLMs to assess a child’s knowledge level and adjust their support accordingly. For example, the LLM often introduced new ideas that required a certain level of understanding, which could be challenging for beginners. Similarly, it should be able to identify incorrect assumptions about the child’s knowledge or skill level and rectify them, ensuring the child is not overwhelmed or misinformed.

Interestingly, the LLM was found to repetitively suggest the same ideas, limiting the scope for creative thinking. Therefore, future iterations of LLM models should aim to generate broader suggestions to encourage diverse creative thoughts. This aligns with the need for LLMs to stimulate creativity by suggesting non-conventional or niche project ideas, broadening the child’s exposure to different concepts and genres.

While the LLM proved helpful in debugging, it sometimes failed to identify specific bugs, such as the local “change lives” bug. This highlights the need for more context awareness in the model to understand the overall goal of the program better. In addition, future models should be designed to express uncertainty when appropriate, enabling children to consider a broader range of possibilities and make more informed decisions about their projects.

Finally, regarding communication style, LLM models should balance their tone to avoid overwhelming young learners with overly enthusiastic responses. Instead, they should foster an environment that encourages exploration and learning at the child’s own pace.

In conclusion, our study underscores the potential of LLMs in enhancing creative coding for children and the need for future iterations to address the highlighted areas of improvement. Our findings contribute to the growing discourse on using LLMs in coding education and align with prior work advocating for culturally-responsive, family-based creative coding contexts. These insights will inform the design of future LLMs, ultimately promoting inclusive and effective computing education.

\subsection{Design Guidelines for AI-Enhanced Creative Coding Tools}
Our current and previous research \cite{Druga2023DesignAI} suggests some guidelines for designing AI-enhanced creative coding tools. The tools should not answer the learners but guide them through their creative process. The fact that in our outputs, the LLM gives options and different suggestions for the youth to evaluate and pick from is a first step in this direction. The LLM should adapt the output to the right learning level based on the youth's age, prior experience, and reactions to its previous suggestion. The LLM’s diversity of answers in our study showcases the possibility of reaching different learning levels, but testing that LLM accurately delivers them is still needed. The following list enumerates more of our design suggestions:

\textit{Promote Agency and Self-expression.} To foster creativity and self-driven learning, LLM tools should stimulate children’s thinking by posing strategic questions instead of providing direct answers \cite{jayagopal2022exploring}. This support style enhances the Scratch coding experience and aligns with the learners’ preference for agency and self-expression.

\textit{Experience Influences Support Needed.} LLM support should be tailored to the coder’s experience level to maximize learning outcomes. For example, novice coders often require more assistance with coding game ideas, whereas intermediate coders may benefit more from ideation and debugging support \cite{jayagopal2022exploring}. 

\textit{Explain the Provenance of Suggestions.} LLM tools should provide transparency about how they generate a suggestion and offer information about the source of the code example. This prevents misconceptions and enhances understanding of the AI’s suggestions, which is particularly useful in Scratch projects \cite{yan2022whygen}.

\textit{Multimodal Debugging Support.} LLM tools should offer visual elements alongside text to clarify complex instructions and aid in locating specific programming blocks, especially given the visual nature of creative coding \cite{mcnutt2023study}. This approach aligns with previous research indicating that augmenting text with visuals provides a more natural coding specification method. 

\textit{Voice Input as “Third Hand.”} Voice input can provide a valuable, hands-free interaction method with the LLM tool, especially when children and parents are collaboratively working on their Scratch project \cite{lin2020zhorai, GitHubNe0:online, Serenade80:online}. However, designing this feature for diverse programmers, including children and parents, requires overcoming challenges such as recognizing children’s speech or foreign accents \cite{kennedy2017child}.

\textit{Live Code Execution} Incorporating \textit{liveness} in the coding platform allows for auto-execution of code, helping users quickly identify non-functioning scripts and offering immediate debugging opportunities. This feature aligns with research on the benefits of immediate feedback in education. It can be especially beneficial in family creative coding scenarios where multiple users may collaborate on a single project \cite{tanimoto2013perspective, guo2013online, kang2017omnicode}.

\textit{Support Diverse Ideas and Projects} LLM tools should encourage various project types, including art projects, story-based experiences, and projects that encourage collaborative mechanics, going beyond mainstream or competitive games. This aspect aligns with the broader goal of fostering creativity and diversity in the Scratch coding environment.

\subsection{Future Work}
In our future work, we primarily aim to delve deeper into the capabilities of LLM Companions in facilitating joint family engagement in creative coding. Understanding how LLM can foster shared learning experiences and promote collaboration among family members is still an open question. This will necessitate evaluating LLM models in multi-turn conversation scenarios involving children and parents, allowing us to comprehend better how LLM can support diverse family learning contexts. We also plan to evaluate LLM models on more complex Scratch programs. This will help us cater to a range of programming competencies and extend the utility of LLM in creative coding. Additionally, we aim to explore the potential of LLM models in providing asynchronous support on multiple projects. Finally, drawing inspiration from research on novice design \cite{dow2010parallel}, we hope to empower young coders to develop better programming skills and foster creativity by exploring multiple ideas before receiving feedback. 

\subsection{Limitations}
While our study has provided valuable insights into the potential of LLM models in enhancing creative coding for families, it is not without its limitations. First, the list of input programs we used for evaluating the LLM model was not exhaustive. The Scratch projects we used were a representative sample, but they do not capture the entire range of programs kids create on Scratch. This vast diversity in creative coding ranges from simple animations to complex games, and our sample may not fully represent this spectrum.

Second, our evaluation scenarios focused primarily on code explanation, debugging, and ideation. While these are critical aspects of creative coding, they do not encompass all possible scenarios kids might need support. There are other areas, such as program design, structuring code, or even specific topics like working with clones and lists in Scratch, where LLM assistance could be beneficial but were not included in our study. Future research should include broader coding scenarios and challenges to assess better LLMs’ potential in supporting creative coding for families.

\section{Conclusion}
This study explored the potential of large language models (LLMs) in enhancing creative coding experiences for families using Scratch. Building upon our previous user studies on AI-Assisted family creative coding, we conducted an extensive evaluation to determine how effectively LLMs could assist in understanding game code, debugging programs, and generating innovative ideas for future creative coding projects. Our research involved meticulously analyzing 120 creative coding support scenarios, incorporating LLMs’ responses with and without practice tasks. In addition, our authors independently assessed each scenario on critical criteria, such as accuracy, pedagogical value, and age-appropriate language.

Our findings revealed that LLMs consistently achieved an impressive success rate of over 80\% across different tasks and evaluation criteria, signifying their considerable potential in supporting family-based creative coding. However, as with any emerging technology, there are areas for refinement and improvement. Our research highlighted the need for more context awareness, diversified suggestions, adaptive communication styles, and improved debugging support in future iterations of LLMs. In conclusion, our research contributes valuable insights into the potential of LLMs in family creative coding. It provides a robust foundation for future research and development in AI-supported coding applications. These findings inform the design of more effective, engaging, and inclusive tools for creative coding education, paving the way for more families to experience the joy and learning opportunities that creative coding can provide.


\bibliographystyle{ACM-Reference-Format}
\bibliography{references}
\end{document}